\documentclass[12pt,letterpaper]{article}         % For preprint
\usepackage{jheppub}
\usepackage{epsfig}
\def\be{\begin{eqnarray}}
\def\ee{\end{eqnarray}}

\newcommand\para{\paragraph{}}

\newcommand{\eqn}[1]{(\ref{#1})}

\def\Dslash{\,\,{\raise.15ex\hbox{/}\mkern-12mu D}}
\def\Dbarslash{\,\,{\raise.15ex\hbox{/}\mkern-12mu {\bar D}}}
\def\delslash{\,\,{\raise.15ex\hbox{/}\mkern-9mu \partial}}
\def\delbarslash{\,\,{\raise.15ex\hbox{/}\mkern-9mu {\bar\partial}}}
\def\pslash{\,\,{\raise.15ex\hbox{/}\mkern-9mu p}}
\def\calDslash{\,\,{\raise.15ex\hbox{/}\mkern-12mu {\cal D}}}

\newcommand{\RN}{Reissner-Nordstr\"om}

\def\lae{\mathrel{\mathop{\smash{\lower .5 ex \hbox{$\stackrel<\sim$}}}}}
\def\lae{\mathrel{\mathop{\smash{\lower .5 ex \hbox{$\stackrel>\sim$}}}}}

\preprint{DAMTP-2015-27}

\title{\large Momentum relaxation from the fluid/gravity correspondence}

\author{Mike Blake} 

\affiliation{ Department of Applied Mathematics and Theoretical Physics,
 University of Cambridge, Cambridge, CB3 OWA, UK}

\emailAdd{m.a.blake@damtp.cam.ac.uk}

\abstract{We provide a hydrodynamical description of a holographic theory with broken translation invariance. We use the fluid/gravity correspondence to systematically obtain both the constitutive relations for the currents and the Ward identity for momentum relaxation in a derivative expansion. Beyond leading order in the strength of momentum relaxation, our results differ from a model previously proposed by Hartnoll et al. As an application of these techniques we consider charge and heat transport in the boundary theory. We derive the low frequency thermoelectric transport coefficients of the holographic theory from the linearised hydrodynamics. }

\begin{document}
\maketitle
%\pdfoutput=1
\pagestyle{plain} \setcounter{page}{1}
\newcounter{bean}
\baselineskip16pt

%\title{\Large Magnetothermoelectric Response from Holography}

%\author[a]{Mike Blake}\author[b]{, Aristomenis Donos} \author[c]{and Nakarin Lohitsiri}

%\affiliation[a]{Department of Applied Mathematics and Theoretical Physics, 
%University of Cambridge, 
%Cambridge, CB3 OWA, UK }
%\affiliation[b]{ Centre for Particle Theory and Department of Mathematical Sciences, 
%Science Laboratories, South Road,  Durham, DH1 3LE, UK }
%\affiliation[c]{Trinity College, University of Cambridge, 
% Cambridge, CB2 1TQ, UK }

%\emailAdd{m.a.blake@damtp.cam.ac.uk}  \emailAdd{aristomenis.donos@durham.ac.uk} \emailAdd{nl313@cam.ac.uk}

%\abstract{In this note we study the effects of a magnetic field on transport using holographic models with broken translational invariance. We show that, after carefully subtracting off non-trivial magnetisation currents, it is possible to express the DC transport currents of the boundary theory in terms of properties of a black hole horizon. This allows us to obtain simple analytic expressions for the electrical, thermoelectric and heat conductivity tensors. Our results apply to both isotropic and anisotropic models, including holographic Q-lattices and to certain theories where translational invariance is broken by linear sources for axions. }

%\begin{document}
%\maketitle
%\pagestyle{plain} \setcounter{page}{1}
%\newcounter{bean}
%\baselineskip16pt

\section{Introduction}

\para{}Perhaps the most useful and widely studied application of holography has been to understanding the transport properties of strongly coupled systems \cite{herzog,nernst}. In general, one is faced with a paucity of techniques to calculate real-time transport in theories that do not admit a quasiparticle description. In contrast, holography provides a simple prescription in which one can extract the linear response coefficients of certain strongly coupled field theories from the classical perturbation equations of black holes. 
\para{} The ultimate hope is that the ability to perform these calculations might yield new physical insights into transport at strong coupling. Particular targets for where such insight is necessary include the transport properties of the strange-metals. It is extremely difficult to reconcile the experimental phenomenology of such materials with a quasiparticle or Drude picture of transport \cite{ong}. As such, it is important to develop techniques to study more general models of heat and charge transport. 
\para{}Motivated by this goal, there has recently been a large amount of progress in calculating the transport properties of holographic models with broken translational invariance \cite{jorge1,jorge2,sandiego,vegh,davison,univdc,lattices,andradewithers,Qlattices,lucasac}. In particular, for a large class of theories it is now possible to obtain analytic expressions for all the DC conductivities in terms of horizon data \cite{univdc,lattices,aristosdc,thermo,amoretti1,newblaise1,newblaise2}. More recently, the low frequency AC conductivity has been calculated in a simple example by exploiting a clever decoupling of the bulk currents \cite{richblaise2}.
%
%\para{} One of the most generic features of these results is that, beyond leading order in the strength of momentum dissipation, the transport coefficients contain an `incoherent' (frequency independent) contribution in addition to a `coherent' (Drude) piece \cite{nernst,richblaise1,richblaise2}. Interestingly, the existence of this `incoherent' piece has been suggested to be phenomenologically relevant in regards to the linear resistivity and the Hall angle anomaly of the strange metals \cite{hartnoll, hall}. 
%
\para{} However, despite this progress, it remains unclear what the underlying physical processes that govern transport in these holographic models are. For translationally invariant theories, a physical understanding of the boundary theory, at least at high temperatures, is provided by relativistic hydrodynamics. This allows a systematic exploration of the physics of the boundary theory in terms of a derivative expansion of the hydrodynamic fields \cite{kovtun, fluidgrav}.
\para{}The defining data of relativistic hydrodynamics are a set of constitutive relations which express the energy-momentum tensor $T^{\mu \nu}$ and electrical current $J^{\mu}$ in terms of a local fluid velocity, charge density and energy density. The dynamics of these fields can then be determined by solving the conservations laws for the currents
\begin{eqnarray}
\partial_{\mu} J^{\mu}  = 0 \nonumber \\
\partial_{\mu} T^{\mu \nu}  = 0
\end{eqnarray}
The connection between the bulk geometry and the fluid-dynamics of the boundary theory has been developed in a beautiful series of works (especially relevant are \cite{hubeny, chargedhydro, erdmenger, minwalla}). In particular, by performing a derivative expansion in the charge density $\rho(x)$, energy density $\epsilon(x)$ and fluid velocity $u^{\mu}(x)$ it is possible to derive both the constitutive relations and the conservation equations of the boundary hydrodynamics. 
\para{}In this paper we use this connection to systematically derive a hydrodynamical description of the simplest holographic model with momentum relaxation. Specifically, we consider boundary theories where translational invariance is broken by turning on linear sources $\phi^{(0)}_{\cal A} = k x_{\cal A}$ for massless scalar fields $\phi_{\cal A}$ in the bulk \cite{andradewithers}. The breaking of translational invariance modifies the conservation equations of relativistic hydrodynamics to
\begin{eqnarray}
\partial_{\mu} J^{\mu} &=& 0 \nonumber \\
\partial_{\mu} T^{\mu \nu} &=& \partial^{\nu} \phi^{(0)}_{\cal A} \langle O_{\cal A} \rangle 
\label{currentequation}
\end{eqnarray}
where the Ward identity for the stress tensor controls how momentum relaxes to equilibrium through scattering off the scalars. By performing a derivative expansion both in the fluid variables and the sources $\partial \phi^{(0)}_{\cal A}$ we can extract the boundary theory hydrodynamics. This allows us to obtain the constitutive relations for $J^{\mu}$, $T^{\mu \nu}$ and $\langle O_{\cal A} \rangle$ up to second order in our expansion. 
\paragraph{}Armed with these constitutive relations we can discuss the linear response of the boundary theory.  At leading order in the derivative expansion our model agrees with a previous proposal of Hartnoll et al \cite{nernst} to include momentum relaxation simply by modifying the conservation law for momentum to 
\be
\partial_{\mu} T^{ \mu x} = -\tau^{-1} T^{0 x} 
\label{hartnollward}
 \ee
where $\tau^{-1}$ is a phenomenological momentum dissipation rate that we calculate. However, beyond leading order, we will see that this proposal is too simple to describe the holographic models. The presence of the scalar field introduces corrections both to the constitutive relations and to \eqn{hartnollward} that are crucial to obtain the correct transport properties. 
\paragraph{}
By carefully taking these factors into account, we calculate the thermoelectric response coefficients of the boundary theory perturbatively in $k$. Our results reproduce the usual formulae expressing the DC transport coefficients in terms of horizon data. In addition, it is straightforward to extract the low-frequency AC response. This decomposes into the sum of a coherent piece, arising from the part of the currents proportional to the fluid velocity, and an incoherent contribution arising from the derivatives of the hydrodynamical fields\footnote{The same decomposition was recently found in a related model using completely different techniques in \cite{richblaise2}.}. In short we have constructed, for the first time, a hydrodynamical description that reproduces the known features of transport in these holographic models. 
\para{}The rest of this paper is organised as follows. In Section~\ref{sec:fluidgravity} we define our bulk model and explain briefly how to apply the fluid/gravity correspondence to discuss momentum relaxation. Our main results are the expressions for the constitutive relations of the energy momentum tensor \eqn{stressconstit}, electrical current \eqn{currentconstit} and scalar expectation values \eqn{secondscalarvev}. In Section~\ref{sec:linearhydro} we linearise the hydrodynamics around equilibrium and calculate the thermoelectric transport coefficients of the boundary theory. Finally we end with a brief discussion of the possible future applications of these techniques in Section~\ref{sec:discussion}. An appendix is provided to describe the technical details of our fluid/gravity calculations.  
\section{The fluid/gravity correspondence} 
\label{sec:fluidgravity}
 \paragraph{}In this section we wish to describe how to apply the fluid/gravity correspondence in the context of the simplest holographic model in which we can study momentum relaxation \cite{fluidgrav}. We will therefore use the Einstein-Maxwell action in 5 bulk dimensions, supplemented by a set of massless scalar fields which will be used to break the translational invariance of the boundary theory\footnote{Note that the fact we are working in 3+1 boundary dimensions is for ease of comparison with the majority of the seminal fluid-gravity literature \cite{hubeny,minwalla,chargedhydro,erdmenger}. We expect that the results presented here will generalise straightforwardly to, for instance, 2+1 dimensional theories.}
\be
 S = \frac{1}{16 \pi G_N}\int \mathrm{d}^5 x \sqrt{-g}\bigg[ R + 12 - F^{MN}F_{MN} - \frac{1}{2} g^{MN} \partial_{M} \phi_{\cal A} \partial_{N} \phi_{\cal A} \bigg ] 
 \label{action}
\ee
Here capital Roman indices $M, N$ refer to the bulk coordinates whilst the index ${\cal A} = 1, 2, 3$ runs over the spatial directions of the boundary theory. The motivation for including several scalar fields is that this allows the construction of bulk solutions which break translational invariance in the boundary but preserve homogeneity and isotropy in the bulk. In particular there exist explicit black hole solutions to \eqn{action} when one applies the sources $\phi_{\cal A}^{(0)} = k x_{\cal A}$ in the boundary. Various aspects of the thermodynamic and linear response coefficients of these solutions have been intensely studied in \cite{andradewithers, richblaise1, richblaise2}. 
\paragraph{}The goal of this work is to provide a hydrodynamical description of the boundary physics dual to \eqn{action}. Our starting point is the familiar translationally invariant \RN\ black brane. If we denote the bulk coordinates by $(r, x^{\mu})$, with $r$ the holographic radial direction, we can write this black hole metric as 
\begin{eqnarray}
g_{M N} dx^{M} dx^{N} &=& - 2 u_{\mu} dx^{\mu} dr - r^2 f(r) u_{\mu} u_{\nu} dx^{\mu} dx^{\nu} + r^2 P_{\mu \nu} dx^{\mu} dx^{\nu} \nonumber \\
A_M dx^{M} &=&  \frac{\sqrt{3} q}{2 r^2} u_{\mu} dx^{\mu} \nonumber \\
\phi_{\cal A} &=& \phi_{\cal A}^0
\label{blackhole}
\end{eqnarray}
where we have introduced a constant 4-velocity aligned in the ingoing-Eddington Finkelstein direction,  i.e. $u_{\mu} dx^{\mu} =-dv$,  and $P_{\mu \nu} = \eta_{\mu \nu} + u_{\mu} u_{\nu}$ projects onto directions perpendicular to $u^{\mu}$. 
\para{}Notice that, because they are massless, we can turn on constant modes of the scalars, $\phi_{\cal A}^0$, without affecting the background geometry. As such, $f(r)$ is just the familiar emblackening factor of the \RN\ black brane
\be
f(r) = 1 - b/r^4 + q^2/r^6
\ee 
which has a horizon at radius $r_0$ satisfying
\be
b = r_0^4\bigg( 1 + \frac{q^2}{r_0^6} \bigg)
\ee
The thermodynamics of the boundary theory is well known. The chemical potential, $\mu$, charge density, $\rho$, energy density $\epsilon$, pressure $P$, temperature, $T$, and entropy density, $s$, are given by 
\be
\mu = \frac{\sqrt{3} q}{2 r_0^2} \;\;\;\;\ \rho = \frac{q}{4 \pi G_N} \;\;\;\;\;\;  \epsilon = \frac{3 b}{ 16 \pi G_N} \;\;\;\;\;\; P = \frac{b}{16 \pi G_N} \;\;\;\;\; T = \frac{r_0}{2 \pi} \bigg(2 - \frac{q^2}{r_0^6} \bigg) \;\;\;\;\;\; s = \frac{r_0^3}{4 G_N} \nonumber
\ee
Finally the expectation values of the scalars are trivial
\be
\langle O_{\cal A} \rangle = 0
\ee
\subsection*{The derivative expansion} 
\label{fluidintro}
\paragraph{}We now wish to construct the hydrodynamics of the boundary theory in a derivative expansion. This correspondence, pioneered in \cite{hubeny}, provides an algorithmic way to construct the hydrodynamics dual to a bulk action and has been subsequently applied to many different models. In particular, the fluid gravity correspondence has been developed in detail for the Einstein-Scalar action in the context of driven fluid flows \cite{minwalla}. Obtaining the hydrodynamics dual to \eqn{action} essentially amounts to generalising these calculations to a charged geometry \cite{chargedhydro,erdmenger}.
 \paragraph{}Whilst the application of the fluid/gravity correspondence is conceptually straightforward it is rather computationally intense. In this paper, our main focus is on understanding the hydrodynamics of the boundary theory. As such, we will be somewhat schematic in presenting the technical details of the gravity calculations. The specific computations we have performed are explained in more detail in Appendix~\ref{appendix1}.
\paragraph{}The starting point of this algorithm is to promote the parameters of the background solution $u^{\mu}, q, b, \phi^0_{\cal A}$ to functions of the boundary coordinates $x^{\alpha}$ to obtain
\be
g^{(0)}_{M N} dx^{M} dx^{N}   &=& - 2 u_{\mu}(x^{\alpha}) dx^{\mu} dr - r^2 f(r, q(x^{\alpha}),b(x^{\alpha})) u_{\mu}(x^{\alpha}) u_{\nu}(x^{\alpha}) dx^{\mu} dx^{\nu} + r^2 P_{\mu \nu}(x^{\alpha}) dx^{\mu} dx^{\nu} \nonumber \\
A_M^{(0)}dx^{M} &=&  \frac{\sqrt{3} q(x^{\alpha})}{2 r^2} u_{\mu}(x^{\alpha}) dx^{\mu} \nonumber \\
\phi_{\cal A}^{(0)} &=& \phi_{\cal A}(x^{\alpha})
\label{ansatz}
\ee
where we can think of $u^{\mu}(x^{\alpha})$ as corresponding to a local fluid velocity in the boundary theory satisfying $u^{\mu} u_{\mu} = -1$. Now in general \eqn{ansatz} is not a solution to the equations of motion of \eqn{action}. Nevertheless, the essential idea is that, because it satisfies the equations of motion on constant solutions, the corrections to \eqn{ansatz} must be proportional to derivatives of the fields $u^{\mu}(x^{\alpha}), q(x^{\alpha}), b(x^{\alpha}), \phi_{\cal A}(x^{\alpha})$. These correction pieces can be determined by solving the Einstein-Maxwell-Scalar equations order by order in this derivative expansion. 
\paragraph{} Before we begin solving for these correction pieces, there are two aspects of our derivative expansion we should highlight. Firstly, we wish to implement momentum relaxation in the boundary theory through position dependent sources for the scalar fields. Just as for constructing black hole solutions, the discussion is simplest if we choose the isotropic sources
\be
\phi_{\cal A}^{(0)} = k x_{\cal A}
\label{sources}
\ee
These sources \eqn{sources} define the laboratory frame of our theory - we will continue to use the calligraphic index ${\cal A}$ to refer to the spatial directions in this frame. 
\paragraph{}Note, however, that the stress tensor of the massless scalars is quadratic in these derivatives. As such, these sources do not influence the metric and gauge field, and hence the boundary constitutive relations, until order $(\partial_{\mu} \phi^{(0)}_{\cal A})^2$ in the derivative expansion. Physically this manifests itself in the fact that at leading order momentum will relax in our models at a rate $\tau^{-1} \sim k^2$. To see the effects of this scattering it is therefore necessary to study fluid flows at frequencies $\omega \sim k^2$.
%\footnote{Physically this is manifest in the fact that at leading order momentum will relax at a rate $\tau^{-1} \sim k^2$. To see the effects of this scattering we therefore need to study fluid flows at frequencies $\omega \sim k^2$.}. 
As a result, we will need to pursue the expansion to higher order in the derivatives of the scalar field than for the other fields. Formally we implement this by taking the scalings
\be
k \sim \varepsilon \;\;\;\;  \partial_{\mu} q \sim \varepsilon^2, \;\;\;\; \partial_{\mu} b \sim \varepsilon^2 \;\;\;\  \partial_{\mu} u^{\alpha} \sim \varepsilon^2
\ee
and then constructing the solutions as a perturbation series\footnote{This expansion can be controlled by fixing the ratio $\mu/T$ and then taking the large $T$ limit. Working perturbatively in derivatives is then valid provided that all fields vary on length scales $l \gg T^{-1}$. } in $\varepsilon$.
\paragraph{}It is worth emphasising that, beyond leading order, the equilibrium configuration is no longer the translationally invariant black brane \eqn{blackhole}. This is because we can turn on position dependent sources for the scalars, $\phi^{(0)}_{\cal A} = k x_{\cal A}$, without inducing a fluid flow. The equilibrium solutions in the presence of these sources can be constructed order by order within our derivative expansion. To do this we simply need to set $u^{\mu} = (1,0,0,0)$ and $ \partial_{\mu} q = \partial_{\mu} b = 0$, but retain the terms that arise from derivatives of the scalar fields. The resulting solutions are equivalent to expanding the black branes of \cite{andradewithers} perturbatively in $k$.
\paragraph{}Secondly, whilst the fluid-gravity correspondence provides us with the full non-linear hydrodynamics, we are ultimately interested in linearising these results around equilibrium. In order to streamline our calculation, we will therefore systematically ignore certain terms that will not appear in the linear theory. In particular, we will neglect terms of the form
\be
u^{\mu} u^{\nu} \partial_{\mu} \phi^{(0)}_{\cal A} \partial_{\nu} \phi^{(0)}_{\cal B}
\ee
For our choice of scalar sources these terms would be quadratic in the fluid velocity and so would play no role in our discussion. Note that in contrast, if we had wanted to study forced fluid flows in which $\partial_{t} \phi^{(0)}_{\cal A}$ was non-zero, then these terms would be important and would need to be determined. 
\paragraph{}With this caveat in mind, our goal is to pursue this derivative expansion to evaluate the constitutive relations for $J^{\mu}$ and $T^{\mu \nu}$ to ${\cal O}(\varepsilon^2)$ and to calculate $\langle O_{\cal A} \rangle$ at ${\cal O}(\varepsilon^3)$. This is enough information to obtain the Ward identities \eqn{currentequation} at ${\cal O}(\varepsilon^4)$ and hence study non-trivial aspects of momentum relaxation in the boundary theory.
\subsection*{Scalar solution at ${\cal O}(\varepsilon)$}
\para{}The first thing we wish to do is to determine the scalar profile at leading order in our expansion. The fluid/gravity correspondence works by solving for the correction to the ansatz \eqn{ansatz} ultra-locally - i.e. at a specific point. More precisely, we use the Lorentz invariance of the zeroth order background to choose coordinates in which we have $u^{\mu} = (1,0,0,0)$ at the point $x^{\mu} = 0$. Note that this new coordinate scheme, in which the fluid is at rest, is different from the laboratory frame defined by the sources $\phi^{(0)}_{\cal A} = k x_{\cal A}$.  To emphasise this we use indices $i=1,2,3$ to refer to the three directions orthogonal to the fluid flow. 
\para{}We can then proceed by writing the scalar field in terms of the ansatz \eqn{ansatz} and a correction piece
\be
\phi_{\cal A}(r, x^{\mu}) =\phi^{(0)}_{\cal A} + \phi^{(1)}_{\cal A} 
\ee
where the correction piece $\phi_{\cal A}^{(1)}$ can be determined by expanding the scalar wave equation 
\be
\partial_{M} \bigg(\sqrt{-g} g^{MN} \partial_N \phi^{(1)}_{\cal A} \bigg) = 0
\ee
up to ${\cal O}(\varepsilon)$ and then solving at the point $x^{\mu} =0$. This results in the differential equation
\be
\partial_r(r^5f(r)\phi^{(1)\prime}_{\cal A}(r)) = -3 r^2 \partial_v \phi^{(0)}_{\cal A}
\ee
which is straightforward to integrate to find the solution
\be
\phi^{(1)}_{\cal A}(r) = \int_{r}^{\infty} \mathrm{d\tilde{r}} \bigg(\frac{\tilde{r}^3 - r_0^3}{\tilde{r}^5 f(\tilde{r})} \bigg) \partial_{v} \phi^{(0)}_{\cal A}
\ee
Note that we have fixed the two constants of integration by demanding regularity at the horizon and normalizability in the UV. From this solution we can extract the expectation value of the operator, $\langle O_{\cal A} \rangle$, from the UV asymptotics as 
\be
-16 \pi G_N \langle O_{\cal A} \rangle = \lim_{r \rightarrow \infty} \sqrt{-g} g^{M r} \partial_{M} \phi_{\cal A} + \frac{1}{2}\partial_{M}(  \sqrt{-\gamma} \gamma^{M N} \partial_{N} \phi_{\cal A} )
\label{scalarvev}
\ee
where $\gamma^{M N}$ is the induced metric on the boundary. At leading order this yields
\be
\langle O_{\cal A} \rangle = -\frac{r_0^3}{16 \pi G_N} \partial_v \phi^{(0)}_{\cal A} = -\frac{r_0^3}{16 \pi G_N} u^{\mu} \partial_{\mu} \phi^{(0)}_{\cal A}
\ee
%
%Note that even at this order in the derivative expansion we can say something about the momentum relaxation of the boundary theory. The Ward identity for momentum relaxation reads
%
%\be
%\partial_{\mu} T^{\mu x} = \partial^{x} \phi  \langle O \rangle
%\ee 
%
%At leading order, we also have the constitutive relations $T^{0 \mu} = (\epsilon = P)u^{\mu}$ and $J^{\mu} = \rho u^{\mu}$ for the momentum density and current. Using $T^{xx} = P$ the Ward identity at leading order becomes
%
%\begin{eqnarray}
%\partial_{0} T^{0 x} &=& \rho \partial^{x} \mu + s \partial^{x} T+ r_0^3 \partial^{x} \phi (u. \partial \phi) \nonumber \\
%&=& \rho \partial^{x} \mu + s \partial^{x} T + T^{0 x}/\tau
%\end{eqnarray} 
%
%where the momentum relaxation rate is given by $\tau^{-1} = r_0^3(\partial_x \phi)^2/(\epsilon + P)$. We therefore see that at leading order you do indeed get the expected momentum relaxation described by Hartnoll's model. However, at next order in the derivative expansion there will be corrections to this result. Our goal is to calculate the leading order corrections to the constitutive relations for the current and energy density, and then finally to the vev of the scalar. 

\subsection*{Metric and gauge field at ${\cal O}(\varepsilon^2)$}
\paragraph{}Our next task is to determine the leading corrections to the metric and gauge field. Once again we pick coordinates so that $u^{\mu} = (1,0,0,0)$ at $x^{\mu}=0$. We can then expand the ansatz \eqn{ansatz} around this point to first order in derivatives to get an expression for the metric
\be
g^{(0)}_{M N} dx^{M} dx^{N} = &2& dv dr - r^2 f(r,q_0,b_0) dv^2 + r^2 dx_{i} dx^{i} - 2 x^{\alpha} \partial_{\alpha} \beta_{i} dx^{i} dr - \nonumber \\ 
&2& x^{\alpha} \partial_{\alpha} \beta_{i} r^2 (1 - f(r,q_0,b_0)) dx^{i} dv - \bigg( \frac{2 q_0 x^{\alpha} \partial_{\alpha} q}{r^4} - \frac{x^{\alpha} \partial_{\alpha} b}{r^2} \bigg) dv^2 \nonumber
\ee
where at this order we can define spatial components of the fluid velocity via $u^{\mu} = (1, \beta_i(x^{\alpha}))$ and we will denote the charge and energy densities at $x^{\mu} = 0$ as $q_0,b_0$. 
\paragraph{}Likewise we can expand the gauge field to get
\be
A^{(0)}_M dx^{M} = -\frac{\sqrt{3}}{2} \bigg[ \bigg( \frac{q_0 + x^{\alpha} \partial_{\alpha} q}{r^2} \bigg) dv - \frac{q_0}{r^2} x^{\alpha} \partial_{\alpha} \beta_i dx^{i}\bigg]
\ee
Just as in the scalar case these fields are not solutions to the equations of motion. Our goal is to find the correction pieces $g_{M N}^{(1)}$, $A_{M}^{(1)}$ so that the full fields
\begin{eqnarray}
g_{M N} &=& g_{M N}^{(0)} + g_{M N}^{(1)} \nonumber \\
A_{M} &=& A^{(0)}_{M} + A^{(1)}_{M}
\end{eqnarray}
satisfy the Einstein-Maxell equations
\be
G_{M N} - 6 g_{M N} + 2 \bigg[ F_{M P} F^{P}_{\;\; N} + \frac{1}{4} g_{M N}F_{P Q} F^{P Q} \bigg ] &=& \frac{1}{2} \bigg[ \partial_M \phi_{\cal A} \partial_N \phi_{\cal A} - \frac{1}{2} g_{MN} \partial_P \phi_{\cal A} \partial^P \phi_{\cal A} \bigg] \nonumber \\
\partial_{M} \bigg( \sqrt{-g} F^{M}_{\;\;\;\; N} \bigg) &=& 0
\label{einsteinmaxwell}
\ee
up to order ${\cal O}(\varepsilon^2)$. Note to obtain the equations at this order, it is necessary to use the solution for the scalar up to ${\cal O}(\varepsilon)$, which we determined in the previous section.
\paragraph{}Conceptually, the resulting equations fall into two classes. Firstly, a subset of these equations are constraints which only act on the boundary data. These are nothing other than equations of motion for the currents \eqn{currentequation} evaluated at ${\cal O}({\varepsilon^2})$ \label{eqncurrent}. These tell us that at any given order in our expansion, the various derivative terms of the hydrodynamics fields are not all independent. Rather charge conservation implies 
\be
\partial_v q = - q_0 \partial_{i} \beta_{i}
\ee
whilst the equation for the stress tensor is equivalent to
\begin{eqnarray}
3 \partial_{v} b &=& - 4 b_0 \partial_{i} \beta_{i} + r_0^3 (\partial_{v} \phi^{(0)}_{A})^2  \nonumber \\
\partial_{i} b &=& - 4 b_0 \partial_{v} \beta_{i} - r_0^3 \partial_{v} \phi^{(0)}_{\cal A} \partial_{i} \phi^{(0)}_{\cal A} 
\label{constraints}
\end{eqnarray}
\paragraph{}The remaining Einstein-Maxwell equations are a set of coupled ordinary different equations that can be used to determine the corrections $g_{MN}^{(1)}$ and $A_{M}^{(1)}$. The calculation is a straightforward generalisation of the analogous computation in the translationally invariant case \cite{chargedhydro,erdmenger}, but with additional terms arising from the scalar stress tensor. The details of the equations and expressions for the resulting metric can be found in Appendix~\ref{appendix1}. 
\paragraph{} Having determined the solution for the metric and the gauge field at leading order, it is straightforward to extract the stress tensor in the boundary from the definition
\be
-\langle T^{\mu \nu} \rangle  = \frac{1}{16 \pi G_N} \lim_{r \rightarrow \infty} r^6 \bigg[ 2(K^{\mu \nu} - \gamma^{\mu \nu} K) + 6 \gamma^{\mu \nu} + \frac{1}{2} (\nabla^{\mu}  \phi_{\cal A} \nabla^{\nu} \phi_{\cal A} - \frac{1}{2} \gamma^{\mu \nu} (\nabla \phi_{\cal A})^2 ) \bigg] \nonumber \\
\label{stresstensorextract}
\ee
where $\gamma$ is the boundary metric, $\nabla$ the associated covariant derivative and $K^{\mu \nu}$ is the extrinsic curvature. Simlarly the current is given by 
\be
\langle J^{\mu} \rangle = \frac{1}{4 \pi G_N} \lim_{r \rightarrow \infty} \sqrt{-g} F^{\mu r} 
\label{currentextract}
\ee
\paragraph{}Note that due to the ambiguities in defining the fluid velocity and hydrodynamical variables out of equilibrium, it is necessary to make a choice of frame in order to obtain a unique expression for these currents. In particular we specify our frame by demanding that charge and energy densities are held constant and given by $b_0, q_0$. We can then define the fluid velocity by demanding that the leading correction to the stress tensor $T^{(1)\mu \nu}$ obeys
\be
u_{\mu} T^{(1) \mu \nu} = 0
\ee
which is commonly referred to as the Landau frame condition. With these specifications we can then read off the constitutive relations. 
\subsection*{Thermodynamics}
\paragraph{}Although our primary motivation for constructing this derivative expansion was to understand the transport properties of these theories, we need to momentarily pause and consider their thermodynamics. Once we turn on sources for the scalar fields, the thermodynamics of our theory receives corrections. To determine these corrections, we set the derivatives of the hydrodynamics fields to zero $\partial_{\mu} q = \partial_{\mu} b = 0$ and set the fluid to rest $u^{\mu} = (1,0,0,0)$ everywhere. 
\paragraph{}In the construction of our bulk solutions, we have been working in an ensemble with fixed charge density and energy density. The expressions for these quantities are therefore unchanged
\be
\epsilon = \frac{3 b}{16 \pi G_N} \;\;\; \rho = \frac{\sqrt{3} q}{4 \pi G_N}
\ee
Conversely the conjugate thermodynamics variables $s, T, \mu$ and $P$ receive corrections that we need to determine.  The origin of these changes is that in the presence of sources for the scalar fields the $g_{vv}(r)$ component of the metric changes. In particular we have that 
\be
g_{vv}(r) = - r^2f(r) + \frac{1}{12}  (\partial_{i} \phi_{\cal A})^2 = -r^2f(r) + \frac{1}{4} k^2
\ee
This extra piece shifts the position of the horizon (defined by $g_{vv}(r_h) = 0$) to 
\be
r_h = r_0 + \frac{k^2}{4 r_0 (4 - 2 Q^2)} + \dots
\ee
where the dots indicate higher order terms in the derivative expansion and $Q= q/r_0^3$. We can see that there are corrections to the entropy density
\be
s = \frac{r_h^3}{4 G_N} = \frac{1}{4 G_N} \bigg(r_0^3 + \frac{3 k^2 r_0}{4(4 - 2 Q^2)}\bigg) + \dots
\ee
as well as to the chemical potential
\be
\mu = \frac{\sqrt{3} q}{2 r_h^2} = \frac{\sqrt{3} q}{2 r_0^2} - \frac{\sqrt{3} q k^2}{4 r_0^4(4 - 2 Q^2)} + \dots
\ee
Similarly we find the temperature is now given by
\be
4 \pi T = -g_{vv}'(r_h) = r_0(4 - 2 Q^2) + \frac{k^2}{2 r_0} \bigg(\frac{7 Q^2 - 2}{4 - 2Q^2} \bigg) + \dots
\ee
and finally pressure is 
\be
P = \mu \rho + s T - \epsilon = \frac{1}{16 \pi G_N} \bigg( b_0 + \frac{1}{2} k^2 r_0^2 \bigg) + \dots
\label{pressure}
\ee
These corrections to the thermodynamics will later be important when we discuss the boundary hydrodynamics beyond leading order in $k$. 
\subsection*{Constitutive relations}
Having extracted the thermodynamics we can now reinstate the other fluctuations $\partial_{\mu} q, \partial_{\mu} b$ and consider a general fluid velocity $u^{\mu}$.  At zeroth order in the derivative expansion, the stress tensor is simply that of a perfect fluid
\be
T^{(0) {\mu \nu}} = \frac{3 b}{16 \pi G_N} u^{\mu} u^{\nu} + \frac{b}{16 \pi G_N} P^{\mu \nu}
\ee
where $P^{\mu \nu} = \eta^{\mu \nu} + u^{\mu} u^{\nu}$ is the projector onto directions perpendicular to the fluid velocity. At ${\cal O}(\varepsilon^2)$, the correction to the stress tensor (in Landau frame) is given by 
\be
T^{(1){\mu \nu}} =  - \frac{2 r_0^3}{16 \pi G_N} \sigma^{\mu \nu} - \frac{r_0^2}{32 \pi G_N} \Phi^{\mu \nu}
\label{stressconstit}
\ee
where the sources appearing above are
\begin{eqnarray}
\sigma^{\mu \nu} &=& P^{\mu \alpha} P^{\nu \beta} \partial_{(\alpha} u_{\beta)} - \frac{1}{3} P^{\mu \nu} \partial_{\alpha}u^{\alpha} \nonumber \\
\Phi^{\mu \nu} &=& P^{\mu \alpha} P^{\nu \beta} \partial_{\alpha} \phi^{(0)}_{\cal A} \partial_{\beta} \phi^{(0)}_{\cal A} - \frac{1}{3} P^{\mu \nu} P^{\alpha \beta} \partial_{\alpha} \phi^{(0)}_{\cal A} \partial_{\beta} \phi^{(0)}_{\cal A} 
\end{eqnarray}
Similarly we can extract the electric current. At leading order this is simply proportional to the fluid velocity but receives subleading corrections proportional to derivatives
\begin{eqnarray}
J^{\mu} = \frac{\sqrt 3 q}{4 \pi G_N} u^{\mu} - \frac{\sqrt{3}}{4 \pi G_N} \frac{M + 1}{4 M r_0} P^{\mu \nu} (\partial_\nu q + 3 q u^{\lambda} \partial_{\lambda} u_{\nu}) - \frac{1}{4 \pi G_N} \frac{3 \sqrt{3}}{8 r_0^2}\frac{q}{M} (u^{\lambda} \partial_{\lambda} {\phi}^{(0)}_{\cal A}) P^{ \mu \nu} \partial_{\nu} \phi^{(0)}_{\cal A} \nonumber \\
\end{eqnarray}
where $M = b/r_0^4$. Finally it will be convenient to note that the constraint equations \eqn{constraints} allow us to eliminate $P^{\mu \nu} u^{\lambda} \partial_{\lambda} u_{\nu}$ in favour of $P^{\mu \nu} \partial_{\nu} b$ and $P^{\mu \nu} \partial_{\nu} \phi^{(0)}_{\cal A}$. We can therefore write the constitutive relation for the current in the form 
\be
J^{\mu} = \frac{\sqrt 3 q}{4 \pi G_N} u^{\mu} - \frac{\sqrt{3}}{4 \pi G_N} \frac{M + 1}{4 M r_0} P^{\mu \nu} (\partial_\nu q - \frac{3 q}{4 b} \partial_{\nu} b) - \frac{1}{4 \pi G_N} \frac{3 \sqrt{3}}{16}\frac{q^3}{b^2} (u^{\lambda} \partial_{\lambda} {\phi}^{(0)}_{\cal A}) P^{ \mu \nu} \partial_{\nu} \phi^{(0)}_{\cal A} \nonumber \\
\label{currentconstit}
\ee
Equation \eqn{currentconstit} is our main result so far. The first term two terms, proportional to the fluid velocity and derivatives of $q$ and $b$, are just the usual constitutive relations of relativistic hydrodynamics. The novel piece is the additional term proportional to the derivatives of the scalar fields. The existence of these subleading corrections in the constitutive relation was not accounted for in \cite{nernst} and will be necessary to describe the transport properties of these holographic theories correctly. 
\subsection*{Scalar solution at ${\cal O}(\varepsilon^3)$}
Our final task is to calculate the expectation value of the scalars at next order in this derivative expansion. The calculation proceeds in entirely the same manner as before. Firstly recall that the solution for the scalar at first order can be written as 
\be
\phi_{\cal A} = \phi^{(0)}_{\cal A} + \phi^{(1)}_{\cal A}
\ee
where we determined the leading correction
\be
\phi^{(1)}_{\cal A}(u^{\mu}, q, b) =  F_1(r, q, b) u^{\mu} \partial_{\mu} \phi_{\cal A}
\ee
with the radial dependence given by the integral
\be
F_1(r, q, b) = \int_{r}^{\infty} {\mathrm d \tilde{r}} \frac{\tilde{r}^3 - r_0^3}{\tilde{r}^5 f(\tilde{r})}
\ee
We now fix coordinates such that $u^{\mu} = (1,0,0,0)$ at $x^{\mu} = 0$ and expand the solution to ${\cal O}(\varepsilon^3)$. Note that expanding $F_1(r, q, b)$ introduces a dependence on the derivatives of the charge and energy densities. We then search for a correction piece $\phi^2_{\cal A}(r)$ such that
\be
\phi_{\cal A} = \phi_{\cal A}^{(0)} + \phi^{(1)}_{\cal A} + \phi^{(2)}_{\cal A}
\ee 
solves the scalar wave equation up to terms of ${\cal O}(\epsilon^3)$. Note that this involves evaluating the wave-equation in the ${\cal O}(\varepsilon^2)$ geometry we determined in the last section. 
\paragraph{}The resulting equation for $\phi^{(2)}_{\cal A}(r)$ takes the form
%\
\be
-(r^5 f(r) \phi^{(2)\prime}_{\cal A}(r))' = S_{\cal A}(r, \partial_i \phi^{(0)}_{\cal B}, \partial_v \phi^{(0)}_{\cal B}, \partial_i q, \partial_i \beta_i, \partial_v \beta_i)
\ee
where an expression for the source term $S_{\cal A}$ can be found in Appendix~\ref{appendix2}. Integrating this equation twice and imposing the appropriate boundary conditions allows us to solve for the scalar fields $\phi^{(2)}_{\cal A}$. From these solutions we can extract the expectation values using \eqn{scalarvev} as
\begin{eqnarray}
\langle O_{\cal A} \rangle =&-& \frac{r_0^3}{16 \pi G_N} u^{\lambda} \partial_{\lambda} \phi^{(0)}_{\cal A} - \frac{\lambda r_0^2}{16 \pi G_N} {\cal S}^+_{\cal A} + \frac{r_0^2}{16 \pi G_N} {\cal S}^-_{\cal A} \nonumber \\ &-& \frac{1}{16 \pi G_N} \frac{r_0}{8 (2 -Q^2)} u^{\lambda} \partial_{\lambda} \phi^{(0)}_{\cal A} (P^{\mu \nu} \partial_{\mu} \phi^{(0)}_{\cal B} \partial_{\nu} \phi^{(0)}_{\cal B}) \nonumber \\ &+& \frac{1}{16 \pi G_N} \frac{3 r_0}{8 M} u^{\lambda} \partial_{\lambda} \phi^{(0)}_{\cal B} (P^{\mu \nu} \partial_{\mu} \phi^{(0)}_{\cal A} \partial_{\nu} \phi^{(0)}_{\cal B}) \nonumber \\ &-&\frac{1}{16 \pi G_N} \frac{6 q P^{\mu \nu} \partial_{\mu} \phi^{(0)}_{\cal A}( \partial_{\nu} q + 3 q u^{\lambda} \partial_{\lambda} u_{\nu})}{24 M r_0^4} 
\label{secondscalarvev}
\end{eqnarray}
where ${\cal S}^{\pm}_{\cal A}$ are related to various derivatives of the fluid velocity
\begin{eqnarray}
{\cal S}^{\pm}_{\cal A} &=& (u^{\lambda} \partial_{\lambda} u^{\mu} \pm \frac{1}{3}\partial_{\lambda}u^{\lambda} u^{\mu}) \partial_{\mu} \phi^{(0)}_{\cal A} 
\end{eqnarray}
and $\lambda$ is a complicated function of the dimensionless variable $Q$. For the neutral theory, $Q=0$, we have that $\lambda = \mathrm{ln}2/2$ but in general we have only been able to calculate it perturbatively (see Appendix~\ref{appendix1}). Fortunately for us, the precise value of $\lambda$ will not be important in understanding the physics of the boundary theory. 
\para{} In principle we could continue performing this $\varepsilon$ expansion to obtain the hydrodynamics of the boundary theory up to any order in the derivative expansion. However, we will not pursue this expansion any further here. Rather, we now have enough information to study momentum relaxation, which we address in detail in the next section. 
\section{Linear response of the boundary theory} 
\label{sec:linearhydro}
\paragraph{} In the last section we explained how to use the technical machinery of the fluid/gravity correspondence to derive the boundary hydrodynamics of our holographic theory in a derivative expansion. We now wish to address a particular physical application of these results, which is to the transport of charge and heat in the boundary QFT. 
\paragraph{}There has recently been a large amount of interest in calculating the thermoelectric transport coefficients of holographic models with broken translational invariance. These coefficients are defined in terms of the linear response of the system to electric fields and thermal gradients as
\[
 \Bigg( \;\; \begin{matrix} J_x \\ Q_x \end{matrix} \;\; \Bigg) =\Bigg( \;\; \begin{matrix} \sigma \;&\; \alpha   \\ 
 T  \alpha \;&\; \bar{\kappa}  
 \end{matrix} \;\; \Bigg) \Bigg( \;\; \begin{matrix} -\partial_{x} \mu \\ -\partial_{x} T \end{matrix} \;\; \Bigg) 
\]
where $Q_x = T^{0x} - \mu J_x$ is the heat current. Then $\sigma$ is just the usual electrical conductivity, $\alpha$ is the thermoelectric conductivity and $\bar{\kappa}$ the heat conductivity. 
\paragraph{}For many holographic models, exact expressions are known for the DC ($\omega = 0$) transport coefficients in terms of horizon data. In particular, for the 5-dimensional Einstein-Maxwell-Scalar model  \eqn{action} with translational invariance broken by the sources $\phi_{\cal A} = k x_{\cal A}$ these read \cite{andradewithers, thermo}
\begin{eqnarray}
\sigma_{DC} &= & \frac{r_0}{4 \pi G_N} + \frac{4 \pi \rho^2}{k^2 s} \nonumber \\
\alpha_{DC} &=& \frac{4 \pi \rho}{k^2} \nonumber \\
\bar{\kappa}_{DC} &=& \frac{4 \pi s T}{k^2}
\label{dccond}
\end{eqnarray}
Importantly these results are exact - that is they are valid for any value of $k$, including when momentum relaxation is very strong. At leading order, that is ${\cal O}(\varepsilon^{-2})$, the holographic formulae agree with the hydrodynamic model studied in \cite{nernst}. In contrast at ${\cal O}(\varepsilon^0)$ there are differences and the results of \cite{nernst} no longer apply. Very recently, a similar story was found to hold for the low-frequency AC conductivity of the 3+1 dimensional analogue of this theory \cite{richblaise2}. 
\paragraph{}Given the discussion so far in this paper, such discrepancies should not come as a surprise. Rather, we have already emphasised that beyond leading order in the $\varepsilon$ expansion our expressions for the constitutive relations for $J^{\mu}, T^{\mu \nu}$ and $\langle O_{\cal A} \rangle$ receive corrections from the scalar fields that were not present in \cite{nernst}. Our goal now is to use to calculate the thermoelectric conductivities up to ${\cal O}(\varepsilon^{0})$ using the hydrodynamical model we derived in Section~\ref{sec:fluidgravity}. We will find that, upon taking into account these corrections, our results precisely agree with the exact holographic formulae. 
\subsection*{Linearised Hydrodynamics}
\paragraph{}The equilibrium configuration of our holographic model is defined by placing the fluid at rest, $u^{\mu} = (1, 0 ,0,0)$, and setting the gradients $\partial_{\mu} q$, and $\partial_{\mu} b$ to vanish. The sources $\phi^{(0)}_{\cal A}$, however, remain non-zero and break the translational invariance of the boundary theory. In Section~\ref{sec:fluidgravity} we computed the relevant thermodynamics of this model up to ${\cal O}(k^2)$. In order to calculate the thermoelectric transport coefficients we need to consider perturbations away from equilibrium
\begin{eqnarray}
\mu(x, t) \rightarrow  \mu + \delta \mu(x, t) \nonumber \\
T(x, t)  \rightarrow T + \delta T(x, t) 
\end{eqnarray}
In this paper we will be only interested in studying the frequency dependence of the transport coefficients and so consider the perturbations $\partial_x \mu$ and $\partial_x T$ to be position (i.e. $x$) independent. We then take a fluid velocity of the form
\be
u^{\mu} = (1, u_x(t), 0, 0) 
\ee
We now proceed to linearise the hydrodynamical model derived in Section~\ref{sec:fluidgravity} in these perturbations $u_x(t), \delta \mu(x,t)$ and $\delta T(x,t)$. 
\paragraph{}After doing this, the constitutive relation for the momentum density $T^{0 x}$ can be written as
\be
T^{0x} &=& \frac{4 b_0}{16 \pi G_N} u_x + .... \nonumber \\
&=& (\epsilon + P) u_x - \frac{k^2 r_0^2}{32 \pi G_N} u_x +..
\ee
where we have used \eqn{pressure} and the $\dots$ indicate terms of higher order in the $\varepsilon$ expansion. The corresponding expression for the electrical current is
\begin{eqnarray}
J_x &=& \frac{\sqrt 3 q}{4 \pi G_N} u_{x} - \frac{\sqrt{3}}{4 \pi G_N} \frac{M + 1}{4 M r_0} (\partial_x q - \frac{3 q}{4 b} \partial_{x} b ) - \frac{1}{4 \pi G_N} \frac{3 \sqrt{3} q^3}{16 b^2}k^2 u_x + ... \nonumber \\
&=& \rho u_x - \sigma_Q \bigg( \partial_{x} \mu - \mu \frac{\partial_x T}{T} \bigg) - \frac{1}{4 \pi G_N} \frac{3 \sqrt{3} q^3}{16 b^2} k^2 u_x + ...
\label{currentcr}
\end{eqnarray}
with $\sigma_Q$ being the `universal conductance' of relativistic hydrodynamics
\be
\sigma_{Q} = \frac{r_0}{4\pi G_N}\bigg( \frac{s T}{\epsilon + P} \bigg)^2
\ee
To obtain the final line in \eqn{currentcr} we have traded the perturbations $ \delta q,  \delta b$ with those of $\delta \mu, \delta T$ using 
\begin{eqnarray}
\delta q &=& \bigg(\frac{\partial q}{\partial \mu} \bigg)_{T} \delta \mu + \bigg(\frac{\partial q}{\partial T}\bigg)_{\mu} \delta T  \nonumber \\
\delta b &=& \bigg(\frac{\partial b}{\partial \mu} \bigg)_{T} \delta \mu + \bigg(\frac{\partial b}{\partial T}\bigg)_{\mu} \delta T  
\end{eqnarray}
\paragraph{}We can now explicitly see that the leading (in $k$) terms in these expressions are simply the usual constitutive relations for the currents within relativistic hydrodynamics. However, the presence of the scalars has introduced extra corrections proportional to the fluid velocity which will enter the transport properties at ${\cal O}(\varepsilon^{0})$. 
\paragraph{}Before proceeding to study momentum relaxation through the Ward identity, it turns out to be extremely convenient to rescale the fluid-velocity in order to simplify these expressions. We therefore define a new fluid velocity $v_x$ via
\be
u_x = \bigg(1 + \frac{k^2 r_0^2}{8 b} \bigg) v_x + \dots
\ee
so that even up to ${\cal O}(\varepsilon^2)$ we can continue to write the momentum density as
\be
T^{0x} = (\epsilon + P) v_x \dots
\ee
The constitutive relation for the current can then be expressed in terms of $v_x$ as
\be
J_x = p v_x - \sigma_{Q}\bigg(\partial_{x} \mu - \mu \frac{\partial_x T}{T} \bigg) + \frac{\mu \sigma_Q k^2}{4 \pi T} v_x + \dots
\label{linearcurrent}
\ee
Finally we need an expression for the heat current $Q_x = T^{0x} - \mu J_x$. This reads
\be
Q_x = s T v_x + \mu \sigma_{Q}\bigg(\partial_{x} \mu - \mu \frac{\partial_x T}{T} \bigg) - \frac{\mu^2 \sigma_Q k^2}{4 \pi T} v_x \dots
\label{linearheat}
\ee
where all the thermodynamic factors appearing in these expressions are accurate up to ${\cal O}(\varepsilon^2)$. Written this way the constitutive relations take a remarkably simple form - the only novel terms are the ${\cal O}({k^2})$ pieces appearing in the electrical and heat currents which can naturally be expressed in terms of thermodynamic quantities. 

\subsection*{Momentum Relaxation}
\paragraph{}The constitutive relations for the currents are not enough to define our model, we also need to supplement them with the equations of motion for the currents
\begin{eqnarray}
\partial_{\mu} J^{\mu} &=& 0 \nonumber \\
\partial_{\mu} T^{\mu \nu} &=& \partial^{\nu} \phi_{\cal A}^{(0)} \langle O_{\cal A} \rangle
\end{eqnarray}
The second of these equations is simply the Ward identity for momentum non-conservation in the presence of the position dependent sources $\phi_{\cal A}^{(0)}$ for the scalars \cite{andradewithers}. The relaxation of our fluid velocity $v_x$ to equilibrium is then described by the $x$ component of the stress-tensor equation of motion. We can evaluate this for our perturbations using the expression \eqn{secondscalarvev} for $\langle O_{\cal A} \rangle$. At order ${\cal O}(\varepsilon^2)$ this equation reads
\be
\partial_{t} T^{0 x} + \partial_{x} P &=&  - \frac{k^2 s}{4 \pi} v_x  + \dots \nonumber \\
&=& - \frac{k^2 s}{4 \pi (\epsilon + P)} T^{0x} + \dots 
\label{wardleading}
\ee
which is precisely the theory studied in \cite{nernst} where the momentum relaxation rate is identified as
\be
\tau^{-1} = \frac{k^2 s}{4 \pi (\epsilon + P)} + \dots
\label{firstscatteringrate}
\ee
However, just as in the constitutive relations for the currents, there are subleading corrections to \eqn{wardleading} that differ from \cite{nernst}. Although our expression for the scalar expectation value \eqn{secondscalarvev} is rather complicated, it simplifies dramatically for linearised perturbations. Evaluating the Ward identity gives at ${\cal O}(\epsilon^4)$ gives the remarkably succinct equation\footnote{In deriving this equation we found it helpful to recall \eqn{pressure} to write $\delta T_{xx} = \delta P - k^2 r_0 \delta r_0/(16 \pi G_N)$ and again make use of the constraints \eqn{constraints}.}  
\be
\partial_{t} T^{0 x} + \partial_{x} P &=&  -\frac{k^2}{4 \pi} \bigg[ s v_x + \frac{\mu \sigma_{Q}}{T}\bigg(\partial_{x} \mu - \frac{\mu}{T} \partial_{x} T \bigg) -  \frac{\mu^2 \sigma_Q k^2}{4 \pi T^2} v_x \bigg]  -  \frac{\lambda k^2 r_0^2}{16 \pi G_N} \partial_{t} v_x + \dots \nonumber  \\
&=& -\frac{k^2}{4 \pi}\frac{Q_x}{T} -\frac{\lambda k^2 r_0^2}{16 \pi G_N} \partial_{t} v_x + \dots
\label{acward}
\ee
where we have noted the quantity in square brackets is nothing other than the entropy current $Q_x/T$. 
\paragraph{} Once we go beyond leading order in the $\varepsilon$-expansion we therefore find that the right hand side of \eqn{acward} is not simply proportional to the momentum density $T^{0x}$. Surprisingly it is the heat current, which differs from the fluid velocity $v_x$ through derivatives of the hydrodynamical fields, that more naturally appears in the equation for momentum relaxation. It is not clear to us whether there is any deep reason why this should be the case, but we will shortly see that it is closely related to the existence of simple formulae  \eqn{dccond} for the DC transport coefficients of this model. 
\subsection*{Thermoelectric response coefficients}
We can now turn to our final question of interest which is to extract the response coefficients from our linearised hydrodynamics. To do this we use the Ward identity \eqn{acward} to get an expression for the fluid velocity in terms of the perturbations $\partial_x \mu$ and $\partial_x T$. Using the thermodynamic relationship $\delta P = s \delta T + \rho \delta \mu$ we find
\begin{eqnarray}
v_x = &-& \frac{1}{1- i \omega \tau} \bigg(\frac{4 \pi \rho}{k^2 s} + \frac{\mu \sigma_Q}{s T} + \frac{\mu^2 \rho \sigma_Q}{s^2 T^2} \bigg)\partial_{x} \mu - \frac{1}{1- i \omega \tau} \frac{4 \pi}{k^2} \partial_{x} T  + \dots
\end{eqnarray}
where the momentum relaxation rate, $\tau^{-1}$ is determined up to ${\cal O}(\varepsilon^4)$ to be
\begin{eqnarray}
\tau^{-1} &=& \frac{s k^2}{4 \pi (\epsilon + P)} \bigg[1  - \frac{\mu^2 \sigma_{Q} k^2}{4 \pi s T^2} - \frac{ \lambda k^2 r_0^2}{16 \pi G_N(\epsilon + P)} + \dots \bigg] 
\label{scatteringrate}
\end{eqnarray}
Since the corrections to $\tau^{-1}$ depend on $\lambda$ they are some non-trivial function of $Q$ that we cannot in general express analytically\footnote{Note that a similar expression for the momentum relaxation rate in the 3+1d analogue of our model was presented in \cite{richblaise2}. We have checked that the lower dimensional equivalent of \eqn{scatteringrate} agrees with their expression.}.
\paragraph{} By inserting \eqn{scatteringrate} into the constitutive relations for the currents we can calculate the thermoelectric transport coefficients perturbatively\footnote{For our perturbations, this is simply the usual Kadanoff-Martin prescription for extracting hydrodynamic correlation functions \cite{nernst, kadanoff}. Note, however, that for spatially modulated perturbations then $\delta \mu(x), \delta T(x)$ have their own dynamics and the computation of transport coefficients is more involved.} in $\omega \sim k^2 \sim \varepsilon^2$. At ${\cal O}(\varepsilon^{-2})$ we have the result
\be
\sigma(\omega) &= & \frac{1}{1 - i \omega \tau} \frac{4 \pi \rho^2}{k^2 s}  \nonumber \\
\alpha(\omega) &=&   \frac{1}{1 - i \omega \tau} \frac{ 4 \pi  \rho}{k^2}\nonumber \\ 
\bar{\kappa}(\omega) &=& \frac{1}{1 - i \omega \tau} \frac{4 \pi s T}{k^2} 
\ee
At this order one simply sees a `coherent' metal, i.e. we have a well-defined Drude peak at low frequencies. As expected these expressions match the results of \cite{nernst} with the momentum relaxation rate \eqn{firstscatteringrate}. 
\para{}However the corrections to the constitutive relations and the Ward identity enter at $O{\cal}(\varepsilon^{0})$ and the transport coefficients are no longer equivalent to those in \cite{nernst}. Rather we find
\begin{eqnarray}
\sigma({\omega}) &=& \sigma_{Q} +  \frac{1}{1 - i \omega \tau} \bigg(\frac{4 \pi \rho^2}{k^2 s} + \sigma_{0} - \sigma_Q \bigg)\nonumber \\
\alpha(\omega) &=& \alpha_Q + \frac{1}{1 - i \omega \tau}  \bigg(\frac{4 \pi \rho}{k^2} - \alpha_Q \bigg)\nonumber \\
\bar{\kappa}(\omega) &=& \bar{\kappa}_Q + \frac{1}{1 - i \omega \tau}\bigg(\frac{4 \pi s T}{k^2} - \bar{\kappa}_Q\bigg)
\label{accond}
\end{eqnarray}
where we have defined $\alpha_Q = - \mu/T \sigma_Q$ and $\bar{\kappa}_Q = \mu^2/T \sigma_Q$ and $\sigma_{0} = r_0/{4 \pi G_N}$. In the DC limit $\omega \rightarrow 0$ these expressions reduce to the exact formulae in terms of horizon data \eqn{dccond}. 
\paragraph{}The AC conductivity can be seen to be composed of two distinct pieces. In addition to the coherent Drude peak, there is now an extra `incoherent' (i.e. frequency independent) contribution.  This incoherent contribution to the transport coefficients simply arises from the parts of the constitutive relations that are proportional to the derivatives $\partial_{x} \mu$, $\partial_{x} T$. Conversely, the coherent piece corresponds to the part of the currents that is proportional to the fluid velocity $v_x$ and so relaxes at the rate $\tau^{-1}$.  
\paragraph{}Reassuringly, an identical structure to \eqn{accond} was recently found in the 3+1 dimensional analogue of \eqn{action} through an explicit computation of the Green's functions of the boundary theory \cite{richblaise2}. The fact that we precisely agree with these results is a strong check on the consistency of our hydrodynamical description. 
\paragraph{}One of the most intriguing consequences of \eqn{accond} is that it reproduces the DC conductivity in a rather non-trivial way. In particular, the ${\cal O}(\varepsilon^{0})$ pieces in \eqn{dccond} arise as a result of rather surprising cancellations between the incoherent and coherent pieces.  It was not clear from the discussion in \cite{richblaise2} why the sum of these pieces should be simple. Our hydrodynamical formulation allows us to see that these cancellations are closely related to the form of momentum relaxation in these models. To see this we should look more carefully at the Ward identity 
\be
\partial_{t} T^{0 x} + \partial_{x} P 
&=& -\frac{k^2}{4 \pi}\frac{Q_x}{T} -\frac{\lambda k^2 r_0^2}{16 \pi G_N} \partial_{t} v_x + \dots
\label{acwardagain}
\ee
In the DC limit, we can ignore the time derivatives in this expression and simply balance the forces. Consequently we must satisfy
\be
Q_x = - \frac{4 \pi T}{k^2} \partial_x P =  - \bigg(\frac{4 \pi \rho T}{k^2}\bigg) \partial_{x} \mu - \bigg( \frac{4 \pi s T}{k^2} \bigg) \partial_{x} T
\ee
from which we can immediately read off the expressions for $\alpha_{DC}$ and $\bar{\kappa}_{DC}$\footnote{A more detailed study of the DC limit in these models was subsequently performed in \cite{maghydro}.}. The simplicity of these DC transport coefficients therefore arises from the fact that it is the heat current, not the momentum, which appears on the right hand side \eqn{acwardagain}. That the finite $\omega$ results are more complicated is then not surprising, since the time derivatives in \eqn{acwardagain} now mix the momentum and heat currents together. As a final observation it is natural to speculate that, since the holographic formulae $\eqn{dccond}$ are exact, an equation like \eqn{acwardagain} involving the heat current should continue to hold at all orders in the derivative expansion. 
\section{Discussion}
\label{sec:discussion}
\para{}In this paper we have used the fluid/gravity correspondence to construct a hydrodynamical description of a simple holographic theory in which translational invariance is broken by scalar fields. We systematically derived the constitutive relations for the boundary current, stress tensor and expectation value of these scalars in a derivative expansion.  Beyond leading order in this expansion, we found that the resulting hydrodynamics was different to that studied in \cite{nernst}. 
\para{}We then looked at charge and heat transport in the boundary theory by linearising this hydrodynamics about equilibrium. This allowed us to calculate the low-frequency thermoelectric response coefficients perturbatively in the strength of momentum relaxation. Our model reproduced the known formulae for the DC conductivities in terms of horizon data. We also found that the AC conductivity could be decomposed into `coherent' and `incoherent' parts in the same manner as was recently seen in \cite{richblaise2}. In summary we have constructed, for the first time, a hydrodynamical description of holographic charge transport with momentum relaxation. 
\para{}Now that we have developed this model, there is a myriad of possible directions for future research. In this paper we have only considered the frequency dependence of the linear response functions. There is much more information contained in the spatially resolved transport coefficients. It would be interesting to develop a detailed understanding of diffusive processes in these theories, analogous to the discussion in \cite{richblaise1}. Secondly, since our description differs from that presented in \cite{nernst}, it is natural to revisit the motivating questions of that work. Through applying the fluid/gravity correspondence in a magnetic field we could construct a theory of magnetotransport \cite{magnetous, magnetoitaly, magnetokorea, magnetoandy} and study the frequency dependence of the Hall angle \cite{hall}. 
\para{}Finally, in this paper we have focused on the simplest holographic model in which we could address the question of momentum relaxation. Nevertheless the techniques developed in Section~\ref{sec:fluidgravity} can be extended to more general settings. In particular, although we expect broadly similar results, the correspondence should be developed for 3+1 dimensional bulk actions \cite{ashok}. It would also be particularly interesting to consider more general, in particular inhomogeneous \cite{jorge1,jorge2,inhomogeneous, lattices}, sources for the scalars and understand how the resulting hydrodynamics differs from that presented here. 
\para{}We hope that developing such hydrodynamical descriptions will provide a physical picture of transport in holography and hence allow for a greater understanding of the potential implications of these models for real-world systems. 

\acknowledgments
I am grateful to Sean Hartnoll, David Tong and especially Aristomenis Donos for many useful conversations and for reading a draft of this work. I also thank Richard Davison and Blaise Gouteraux for discussing the results of \cite{richblaise2} with me prior to their publication. I am funded by a Junior Research Fellowship at Churchill College, Cambridge. During the completion of this work I received hospitalilty from the Stanford Insititute for Theoretical Physics and the Galileo Galilei Institute for Theoretical Physics as well as partial support from the INFN. This work was supported in part by the European Research Council under the European UnionÕs Seventh Framework Programme (FP7/2007-2013), ERC Grant agreement STG 279943, Strongly Coupled Systems.

\appendix
\section{Details of the fluid/gravity calculations}
\label{appendix1}
In this appendix we provide, for the benefit of the interested reader, some further details about the fluid/gravity computations we used to obtain the constitutive relations presented in Section~\ref{sec:fluidgravity}. We have already explained, in the main text, how to compute the scalar solution to ${\cal O}(\varepsilon)$ and so begin with the leading corrections to the metric and gauge field. 
\subsection*{Metric and gauge field at first order}
Our goal in this section is to solve the Einstein-Maxwell equations  \eqn{einsteinmaxwell} at $x^{\mu} =  0$ to determine the ${\cal O}(\varepsilon^2)$ corrections to the the metric and gauge field $g^{(1)}_{MN}, A_{M}^{(1)}$. As is standard in these calculations, we proceed by decomposing these equations into their scalar, vector and tensor pieces. 
\subsection*{Scalar modes}
We begin, following the discussion in \cite{chargedhydro}, by parameterising the scalar parts of the metric and gauge field $g_{M N}^{(1)}$, $A_{M}^{(1)}$ as 
\begin{eqnarray}
g^{(1)}_{ii} &=& r^2 h_1(r) \;\;\;\;\;\;\;\; \text{(no sum)} \nonumber   \\
g^{(1)}_{vv} &=& k_1(r)/r^2 \nonumber \\
g^{(1)}_{v r} &=& - \frac{3}{2} h_1(r) \nonumber \\
A_v^{(1)} &=& - \frac{\sqrt{3} w_1(r) }{2 r^2}
\end{eqnarray}
where $g_{ii}$ and $g_{v r}$ are related by a choice of gauge
\be
\mathrm{Tr} \bigg[(g^{(0)})^{-1} g^{(1)}\bigg] = 0 \;\;\;\; g_{rr} = 0 \;\;\;\; g_{r\mu} \propto u_{\mu} \;\;\;\; A_r = 0
\ee
%
%\subsubsection*{Constraint equations}
%
\paragraph{}As we mentioned in the main text, the Einstein-Maxwell equations can be decomposed into constraint equations and dynamical equations. The constraint equations are defined as those that arise from the dot product of the Einstein-Maxwell equations with the vector dual of $dr$.
\paragraph{}Importantly, a subset of these constraint equations correspond to the hydrodynamical equations of motion of the boundary theory. The first such constraint equation, arising from the Maxwell equations $M_N = 0$ is given by
\be
g^{rr} M_{r} + g^{r v} M_v = 0
\ee
which evaluates to the requirement that we have
\be
\partial_{v} q = - q_{0} \partial_{i} \beta_{i}
\ee
and is equivalent to the equation for conservation of the boundary current 
\be
\partial_{\mu} J^{\mu}  = 0
\ee
Similarly we have the constraint equations arising from the Einstein equations $E_{M N}  = 0$. Firstly we can take
\be
g^{rr} E_{v r} + g^{v r} E_{v v} = 0
\ee
which reads
\be
3 \partial_{v} b = - 4 b_0 \partial_{i} \beta_{i} + r_0^3 (\partial_{v} \phi^{(0)}_{\cal A})^2
\ee
and corresponds to the scalar component of the Ward-identity for momentum relaxation
\be
u_{\mu} \partial_{\nu} T^{\mu \nu} = u_{\mu} \partial^{\mu} \phi^{(0)}_{\cal A} \langle O_{\cal A} \rangle
\label{ward}
\ee
Finally there is one more constraint equation arising from
\be
g^{rr} E_{rr} + g^{v r} E_{v r} = 0
\ee
This equation does not act on the boundary data, but rather imposes a relationship between $h_1(r)$, $k_1(r)$ and $w_1(r)$
\begin{eqnarray}
2 \partial_i \beta_i r^5 + 12 r^6 h_1(r) + 4 q_0 w_1(r)  - b_0 r^3 h_1'(r) + 3 r^7 h_1'(r) - r^3 k_1'(r) - 2 q_0 r w_1'(r) = \nonumber \\
- \frac{1}{6} r^4 (\partial_i \phi^{(0)}_{\cal A})^2 + \frac{1}{6} r^8 f(r) (\phi^{(1)\prime}_{\cal A}(r))^2 \nonumber \\
\label{einsteinconstraint}
\end{eqnarray}
%
%\subsubsection*{Dynamical Equations}
%
Together with these constraint equations we also have the remaining dynamical Einstein-Maxwell equations. In particular the $rr$-component of the Einstein equation reads
\be
3 h_1''(r) + 15 h_1'(r)/r = -\phi^{(1)\prime}_{\cal A}(r)^2 
\label{einsteindynamical}
\ee
whilst the $v$-component of the Maxwell equation is
\be
r w_1''(r) - w_1'(r) - 6 q_0 h_1'(r) = 0
\label{maxwelldynamical}
\ee
These three equations (\eqn{einsteinconstraint} \eqn{einsteindynamical} and  \eqn{maxwelldynamical}) can now be solved for the undetermined functions $h_1(r), k_1(r), w_1(r)$. 
\paragraph{} In general solving these coupled equations results in rather ugly expressions for the metric. However, the origin of many of these complications are the terms proportional to $\phi^{(1)\prime}_{\cal A}(r)^2 \sim (u^{\mu} \partial_{\mu} \phi^{(0)}_{\cal A})^2$. As we described in Section~\ref{fluidintro}, these will not contribute to the the linearised hydrodynamics with our choice of sources \eqn{sources}. In order to make our discussion more presentable, we will therefore neglect these terms in equations \eqn{einsteinconstraint} and \eqn{einsteindynamical}.
\paragraph{} If we do this we can then first solve for $h_1(r)$ using \eqn{einsteindynamical} to get
\be
h_1(r) = C_1/r^4 + C_2
\ee
Plugging this into \eqn{maxwelldynamical} yields $w_1(r)$
\be
w_1(r) = C_3 r^2 + C_4 - q_0 C_1/r^4
\ee
Finally, we can then use \eqn{einsteinconstraint} to solve for $k_1(r)$
\be
k_1(r) = \frac{2}{3} r^3 \partial_i \beta_i + \frac{1}{12} r^2 (\partial_i \phi^{(0)}_{\cal A})^2 + C_5 - 2 q_0/r^2 C_4 + (2 q_0^2/r^6 - b_0/r^4) C_1
\ee
All that remains is to fix the various constants of integration that appear in the solution. Firstly, the constants $C_2$ and $C_3$ correspond to non-normalizable modes and so can be 
set to zero in the UV. Secondly, the constants $C_4$ and $C_5$ can be absorbed into the definition of charge and energy density and can also be ignored. Finally the constant $C_1$ can be
removed by a redefinition of the radial coordinate $r \rightarrow r(1+ A/r^4)$. As a result, we can conclude that the solution in the scalar sector is
\be
h_1(r) = w_1(r) = 0 \;\;\;\; k_1(r) = \frac{2}{3} r^3 \partial_{i} \beta_{i} + \frac{1}{12} r^2 (\partial_i \phi^{(0)}_{\cal A})^2
\ee 
where once again we emphasise that in general there are terms proportional to $(\partial_v \phi_{\cal A}^{(0)})^2$ in these functions which we are neglecting for the sake of simplicity. 
\subsubsection*{Vector modes}
We can now proceed to calculate the corrections to the vector channel, that is we wish to solve for the corrections $g^{(1)}_{vi}, A^{(1)}_{i}$. This will be particularly
important for us because it is this calculation that determines the constitutive relation for the electrical current. In this sector we have one constraint equation, arising from 
\be
g^{rr} E_{r i} + g^{r v} E_{vi} = 0
\ee
which gives
\be
\partial_{i} b = -4 b_0 \partial_{v} \beta_{i} - r_0^3 \partial_{v} \phi^{(0)}_{\cal A} \partial_{i} \phi^{(0)}_{\cal A}
\label{vecconstraint}
\ee
and is equivalent to the vector component of the Ward identity for momentum relaxation. The corrections to the metric and gauge field are determined by the dynamical equations. In particular the Einstein equations $E_{r i}$ read
\be
r^2(r g^{(1)\prime}_{vi})' - 4 r g^{(1)}_{v i} + 4 \sqrt{3} q_0 a^{(1)\prime}_i = -3 r^2 \partial_{v} \beta_{i} + r^3 \phi^{(1)\prime}_{\cal A}(r) \partial_{i} \phi^{(0)}_{\cal A}
\label{einsteinvector}
\ee
whilst the integrated Maxwell equations yield
\be
\bigg( r^3 f(r) a^{(1)\prime}_i + \frac{\sqrt{3} q_0 g^{(1)}_{v i}}{r^2} + \frac{\sqrt{3}}{2 r} (q_0 \partial_v \beta_i + \partial_i q) \bigg) = D^1_i
\label{maxwellvector}
\ee
where $D^{1}_i$ is a constant of integration that we will eventually determine by requiring regularity at the horizon. These equations can be solved by using \eqn{einsteinvector} to write
\be
a_i^{(1)\prime} = \frac{1}{4 \sqrt{3} q_0} (- 3 r^2 \partial_v \beta_i + r^3 \phi^{(1)\prime}_{\cal A}(r) \partial_{i} \phi^{(0)}_{\cal A} + 4 r g^{(1)}_{v i} - r^2 (r g^{(1)\prime}_{v i})' )
\ee
and inserting this into the Maxwell equation to get a second order differential equation for $g^{(1)}_{vi}$
\be
r^3 f(r)(4 r g^{(1)}_{vi} - r^2(r g^{(1)\prime}_{vi})') + 12 q_0^2 g^{(1)}_{vi}/r^2 = &-& 6 q_0( \partial_i q+ 3 q_0 \partial_{v} \beta_{i})/r\nonumber  \\  &+& 4 \sqrt{3} q_0 D^1_i - r^6 f(r) \phi^{(1)\prime}_{\cal A}(r) \partial_{i} \phi^{(0)}_{\cal A} \nonumber \\ &+& (3 r^5 f(r) + 12 q_0^2/r) \partial_{v} \beta_i 
\ee
which has the solution
\begin{eqnarray}
g_{v i}^{(1)} &=& \partial_{v} \beta_{i} r + h_{v i}^{(1)} \nonumber \\
h_{vi}^{(1)} &=&  r^2 f(r) \bigg[ \int_r^{\infty} d{\tilde{r}} \frac{1}{f(\tilde{r})^2} \bigg ( \frac{A_i}{3 \tilde{r}^8} - \frac{2 \sqrt{3} q_0 D^1_i}{\tilde{r}^7} + \bigg({\frac{r_0^3}{\tilde{r}^6} + \frac{1}{2 \tilde{r}^3}}\bigg) \partial_{v} \phi^{(0)}_{\cal A} \partial_{i} \phi^{(0)}_{\cal A} + \frac{D_i^2}{\tilde{r}^5} \bigg ) + D^3_i  \bigg] \nonumber 
\label{gvisolution}
\end{eqnarray}
where $A_i = 6 q_0(\partial_i q + 3 q_0 \partial_v \beta_i)$. 
\paragraph{}Once again we need to fix some constants of integration. $D_i^3$ corresponds to a non-normalizable mode and hence should be set to zero. Upon calculating the stress tensor of the dual theory one finds that a non-zero $D_i^2$ gives a contribution to the momentum $T^{(1)}_{0i} \propto D_i^2$. In Landau frame $u^{\mu}T^{(1)}_{\mu i} = 0$ this means we must set $D^2_{i} = 0$. Finally $D^1_{i}$ is fixed by requiring regularity of $a_i^{(1)}$ at the horizon. From \eqn{maxwellvector} this requires that
\be
\frac{\sqrt{3} q_0 g^{(1)}_{v i}}{r^2} + \frac{\sqrt{3}}{2 r} (q_0 \partial_v \beta_i + \partial_i q) -D^{1}_{i} = 0
\ee
at the horizon. Using the solution \eqn{gvisolution} we find that this constraint implies that we take
\be
D^1_i = \frac{\sqrt{3}}{4 r_0} \bigg( \frac{M_0 + 1}{M_0}  \bigg) (\partial_i q + 3 q_0 \partial_{v} \beta_{i}) + \frac{3 \sqrt{3}}{8 r_0^2} \frac{q_0}{M_0} \partial_v \phi^{(0)}_{\cal A} \partial_{i} \phi^{(0)}_{\cal A}
\ee
where $M_0 = b_0/r_0^4$ is the dimensionless mass of the black brane. This determines the metric correction to be 
\begin{eqnarray}
g_{v i}^{(1)} = \partial_{v} \beta_{i} r + r^2 f(r)  \int_r^{\infty} d{\tilde{r}}  \frac{1}{f(\tilde{r})^2} \bigg[&&\bigg ( \frac{1}{3 \tilde{r}^8} - \frac{M_0+1}{4 M_0 r_0 \tilde{r}^7}\bigg)6 q_0(\partial_i q + 3 q_0 \partial_v \beta_i)\nonumber \\  &+& \bigg({\frac{r_0^3}{\tilde{r}^6} + \frac{1}{2 \tilde{r}^3}} - \frac{9 q_0^2}{4 M_0 r_0^2 \tilde{r}^7} \bigg) \partial_{v} \phi^{(0)}_{\cal A} \partial_{i} \phi^{(0)}_{\cal A}  \bigg]  \nonumber \\
\label{gvisolution}
\end{eqnarray}
and we will leave the gauge field as implicitly determined by this result through \eqn{maxwellvector} and the requirement of normalizability in the UV. 
\subsubsection*{Tensor modes}
Finally we can determine the solution in the tensor sector. Here we parameterise the tensor components of the metric via $g_{ij} = r^2 \alpha_{i j}$, where $\alpha_{ij}$ is a tensor of $SO(3)$. In this sector we only have a dynamical Einstein equation which reads
\be
(r^5 f(r) \alpha_{ij}')' = - r \phi_{ij} - 6r^2 \sigma_{ij} 
\ee
where we have defined the source terms
\begin{eqnarray}
\phi_{ij} &=& \partial_i \phi^{(0)}_{\cal A} \partial_j \phi^{(0)}_{\cal A} - \frac{1}{3} \delta_{i j} (\partial_k \phi^{(0)}_{\cal A})^2 \\
\sigma_{ij} &=& \frac{\partial_i \beta_j + \partial_j \beta_i}{2} - \frac{1}{3} \delta_{ij} (\partial_k \beta_k)
\end{eqnarray}
Integrating this equation and imposing regularity and normalizability gives the solution
\be
\alpha_{ij} = - \frac{1}{2} \phi_{i j} \int^{\infty}_{r} \mathrm{d\tilde{r}} \frac{r_0^2 - \tilde{r}^2}{f(\tilde{r}) \tilde{r}^5} - 2 \sigma_{ij} \int^{\infty}_{r} \mathrm{d\tilde{r}} \frac{r_0^3 - \tilde{r}^3}{f(\tilde{r}) \tilde{r}^5}
\ee
which completes the determination of the metric. The expectation values of the stress tensor and current at $x^{\mu}$ can then be read off from equations \eqn{stresstensorextract} and \eqn{currentextract}. Covariantizing these expressions in terms of the fluid dynamical variables $q(x), b(x)$ and $u^{\mu}(x)$ gives the constitutive relations presented in the main text. 
\subsection*{Scalar field at second order}
\label{appendix2}
\paragraph{}We can now present the details of how we determine the scalar field solution to ${\cal O}(\varepsilon^3)$. Expanding the scalar wave equation about the point $x^{\mu}=0$ and using the metric determined in the last section yields the following differential equation for the correction\footnote{Note that the ignored pieces $\sim (\partial_{v} \phi^{(0)}_{\cal A})^2$ in $k_1(r)$ and $h_1(r)$ would contribute additional terms to this equation which we are continue to neglect.} $\phi^{(2)}_{\cal A}(r)$\
\begin{eqnarray}
-(r^5 f(r) \phi^{(2)\prime}_{\cal A}(r))' = (r + 2 r^3 F_1'(r) + 3 r^2 F_1(r)) S^+_{\cal A} - 2 r S^-_{\cal A} \nonumber \\ + F_2'(r) \partial_{v} \phi^{(0)}_{\cal A} (\partial_{i} \phi^{(0)}_{\cal B})^2 - (r h^{(1)}_{v i})' \partial_i \phi^{(0)}_{\cal A} \nonumber \\
\label{scalar2}
\end{eqnarray}
where $F_1(r)$ is the radial profile appearing in the solution for the scalars at first order 
\be
F_1(r) = \int_{r}^{\infty} {\mathrm d \tilde{r}} \frac{\tilde{r}^3 - r_0^3}{\tilde{r}^5 f(\tilde{r})}
\ee
and $F_2(r)$ is defined by
\be
F_2(r) = \frac{ r^4 r_0^2(r^2 + r r_0 + r_0^2)}{12 (r + r_0)(-q_0^2 + r^2 r_0^2(r^2 + r_0^2))}
\ee
The source terms $S^{\pm}_{\cal A}$ are related to derivatives of the fluid velocity 
\begin{eqnarray}
S^{\pm}_{\cal A} &=& \partial_{v} \beta_{i} \partial_{i} \phi^{(0)}_{\cal A} \pm \frac{1}{3}\partial_{i} \beta_i \partial_{v} \phi^{(0)}_{\cal A} 
\end{eqnarray}
and we note that in deriving \eqn{scalar2} we have chosen to eliminate any terms proportional to $\partial_{i} b$ using the constraint \eqn{vecconstraint}. It is straightforward to integrate \eqn{scalar2} to obtain
\be
-(r^5 f(r) \phi^{(2)\prime}_{\cal A}(r)) = &&\int_{r_0}^{r} \mathrm{d\tilde{r}} (\tilde{r} + 2 \tilde{r}^3 F_1'(\tilde{r}) + 3 \tilde{r}^3 F_1(\tilde{r})) S^+_{\cal A} - \int_{r_0}^{r} \mathrm{d \tilde{r}} 2 \tilde{r} S^-_{\cal A}  \nonumber \\&+& F_2(r) \partial_{v} \phi^{(0)}_{\cal A} (\partial_{i} \phi^{(0)}_{\cal B})^2 - r h^{(1)}_{v i} \partial_i \phi^{(0)}_{\cal A} - E_{\cal A}^{(1)} \nonumber \\
\ee
where the constant of integration $E_{\cal A}^{(1)}$ is fixed by demanding regularity at the horizon to be 
\be
E_{\cal A}^{(1)} = \frac{r_0}{8(2 - Q_0^2)} \partial_{v} \phi^{(0)}_{\cal A} (\partial_{i} \phi^{(0)}_{\cal B})^2  - \frac{3 r_0}{8 M_0} \partial_i \phi^{(0)}_{\cal A} (\partial_{v} \phi^{(0)}_{\cal B} \partial_{i} \phi^{(0)}_{\cal B}) + \frac{A_i \partial_{i} \phi^{(0)}_{\cal A}}{24 M_0 r_0^4}
\ee
with $Q_0 = q_0/r_0^3$. The full solution for $\phi^{(2)}_{\cal A}(r)$ can then be obtained by a further integration and requiring normalizability in the UV. Finally we can extract the $\langle O_{\cal A} \rangle$  at $x^{\mu} =  0$ as
\begin{eqnarray}
16 \pi G_N \langle O_{\cal A} \rangle &=& -r_0^3 \partial_{v} \phi^{(0)}_{\cal A} - \lambda r_0^2 S^+_{\cal A} + r_0^2 S^-_{\cal A} - E_{\cal A}^{(1)} \nonumber \\
&=& - r_0^3 \partial_{v} \phi^{(0)}_{\cal A} - \lambda r_0^2 S^+_{\cal A} + r_0^2 S^-_{\cal A}  \nonumber \\ &-&  \frac{r_0}{8(2-Q_0^2)} \partial_{v} \phi^{(0)}_{\cal A} (\partial_{i} \phi^{(0)}_{\cal B})^2 \nonumber \\  &+& \frac{3 r_0}{8 M_0} \partial_i \phi^{(0)}_{\cal A} (\partial_{v} \phi^{(0)}_{\cal B} \partial_{i} \phi^{(0)}_{\cal B})  \nonumber \\ &-& \frac{6 q_0 \partial_{i} \phi^{(0)}_{\cal A}( \partial_i q + 3 q_0 \partial_{v}  \beta_i)}{24 M_0 r_0^4} 
\end{eqnarray}
Covariantizing this expression gives the constitutive relation \eqn{secondscalarvev}. As we mentioned in the main text we have not been able to obtain an exact expression for $\lambda$. It is possible, although not very illuminating, to calculate $\lambda$ perturbatively, for which we find the leading terms
\be
\lambda =  \frac{\mathrm{ln}2}{2} + (3 - \mathrm{ln}4)\frac{Q^2}{4} + \dots
\ee
Fortunately, the exact value of $\lambda$ did not play an important role in our discussion of transport in Section~\ref{sec:linearhydro}.   

\end{document}